\documentclass[twoside,leqno,twocolumn]{article} 
\usepackage{ltexpprt}

\begin{document}

\newcommand{\cmax}{C_{\max}}
\newcommand{\cmaxopt}{C^{{\rm opt}}_{\max}}
\newcommand{\cavg}{C_{\mbox{avg}}}

\newcommand{\qed}{\rule{7pt}{7pt}}



\catcode`@=11

\def\imparray{\stepcounter{equation}\let\@currentlabel=\theequation
\global\@eqnswtrue
\global\@eqcnt\z@\tabskip\@centering\let\\=\@eqncr
$$\halign to \displaywidth\bgroup\llap{${##}$\hskip 4\arraycolsep}\tabskip\z@&
  \@eqnsel\hskip\@centering
  $\displaystyle\tabskip\z@{##}$&\global\@eqcnt\@ne 
  \hskip 2\arraycolsep \hfil${##}$\hfil
  &\global\@eqcnt\tw@ \hskip 2\arraycolsep $\displaystyle\tabskip\z@{##}$\hfil 
   \tabskip\@centering&\llap{##}\tabskip\z@\cr}

\def\endimparray{\@@eqncr\egroup
      \global\advance\c@equation\m@ne$$\global\@ignoretrue}

\@namedef{imparray*}{\def\@eqncr{\nonumber\@seqncr}\imparray}
\@namedef{endimparray*}{\nonumber\endimparray}

\catcode`@=12

\newcommand{\fopt}{f_{\mbox{\tiny opt}}}
\newcommand{\copt}{c_{\mbox{\tiny opt}}}
\newcommand{\aopt}{\alpha_{\mbox{\tiny opt}}}
\newcommand{\rhom}{{\rho}^{-}}
\newcommand{\rhop}{{\rho}^{+}}
\newcommand{\Lra}{\Leftrightarrow}



\title{Improved Bicriteria Existence Theorems for Scheduling}

\author{Javed Aslam\thanks{\protect\scriptsize {\tt
      jaa@dartmouth.edu}. Research partially supported by Walter and
    Constance Burke Research Initiation Award 424879.}
  \and April Rasala\thanks{\protect\scriptsize {\tt
      apes@dartmouth.edu}. }\and
Cliff Stein\thanks{\protect\scriptsize
{\tt cliff@cs.dartmouth.edu}. Research partially supported by NSF
Career Award CCR-9624828.} 
        \and
Neal Young\thanks{\protect\scriptsize {\tt
      ney@dartmouth.edu}. Research partially funded by NSF Career
      Award CCR-9720664.}} 

\date{Dartmouth College}

\maketitle

\pagestyle{myheadings}
\markboth{}{} 


Two common objectives for evaluating a schedule are the {\em makespan}, or
schedule length, and the {\em average completion time}.  In this note,
we give improved bounds on the existence of schedules that
simultaneously optimize both criteria.

In a scheduling problem,
we are given $n$ jobs and $m$ machines.
With each job $j$ we associate a non-negative weight
$w_j.$ A {\em schedule} is an assignment
of jobs to machines over time, and yields a completion time $C_j$ for
each job $j$. We then define the average completion time
as $\sum_{j=1}^n w_j C_j$ and the makespan as $\cmax = \max_j C_j$.  We
use $\cmaxopt$ and $\sum w_jC_j^*$ to denote the optimal makespan and
average completion time.

We will give results which will hold for a wide variety of combinatorial
scheduling problems.  In 
particular, we require that valid schedules for 
the problem satisfy two very general conditions.
First, if we take a valid schedule $S$
and remove from it all jobs that complete after time $t$,
the schedule remains a valid schedule for those jobs that remain.
Second, given two valid schedules 
$S_1$ and $S_2$ for two sets $J_1$ and $J_2$ of jobs (where $J_1 \cap J_2$
is potentially nonempty),
the composition of $S_1$ and $S_2$, obtained by 
appending $S_2$ to the end of $S_1$, and removing from $S_2$ all jobs
that are in $J_1 \cap J_2$, is a valid schedule for $J_1 \cup J_2$.

For
the rest of this note we will make claims about ``any'' scheduling
problem, and mean any problem that satisfies the two  conditions
above.  In addition, if a schedule has 
$\cmax \leq \alpha \cmaxopt$ and $\sum w_jC^j \leq \beta \sum w_jC_j^*$
we call $S$ an   $(\alpha, \beta)$-schedule. 

Stein and Wein \cite{SteinW97} recently gave a powerful but simple
theorem on the existence of schedules which are simultaneously good
approximations for makespan and for average completion time.  
They showed that for {\em any} scheduling problem, there exists a
$(2,2)$-schedule. The construction is simple.   We take an optimal
average completion time schedule and replace the subset $J'$ of jobs
that finish after time $\cmaxopt$ by an optimal makespan schedule for $J'$.  The
schedule has length at most $2\cmaxopt$, and the completion time of each job
at most doubles, thus we obtain a (2,2)-schedule.  

In the (2,2)-schedule, $\cmaxopt$ was the {\em break-point}, the point
at which we truncated the average completion time schedule and started
the makespan schedule on the remaining jobs.  By considering several different break points
simultaneously, and taking the best one, Stein and Wein show, via a
complicated case analysis,  how to
achieve improved approximations.  In particular, they prove the
existence of $(2,1.735)$-schedules, $(1.785,2)$-schedules and
$(1.88,1.88)$-schedules.

In this paper, we give improved theorems on the existence of
bicriteria schedules. Our first conceptual idea is that an average
completion time schedule, appropriately normalized, can be viewed as a
continuous probability density function.  Even though schedules are
actually discrete functions, this mapping to continuous functions
facilitates the analysis.  We choose, for any average
completion time schedule (pdf), the breakpoint that gives the best
bicriteria result; this calculation is now expressed as an integral.
Choosing the pdf that maximizes this integral provides a
worst-case schedule. 

We now give an overview of the technical details. 
Wlog, we can normalize the weights $w_j$ in the optimal average
completion time schedule so that $\sum_j
w_j C_j^* = 1$.  Now let $g(z) = (\sum_{j| C_j^*=z} w_j C^*_j)
\delta(0)$, where $\delta(\cdot)$ is Dirac's delta function. 
By our normalization assumption, we have that
$\int_0^{\infty} g(z)\,dz = 1$ and $g(z) \geq 0$.  Thus $g$ is a
probability density function (pdf).  Let $L$ denote the optimal makespan
and consider the schedule formed by having a
breakpoint at $\alpha L$.  The jobs that complete before time $\alpha
L$ have their completion times unaffected, while those that complete
at time $z > \alpha L$ have their completion times multiplied by
at most $(1 + \alpha)/z$.  Thus, the resulting schedule has a makespan of
$(1+\alpha)L$ and an average completion time of
$$ \int_0^{\alpha L} g(z)\,dz + \int_{\alpha L}^{\infty}
\frac{(1+\alpha)L}{z} g(z)\,dz$$
$$= \int_0^{\infty} g(z)\,dz + \int_{\alpha
L}^{\infty} \frac{(1 + \alpha)L -z}{z} g(z)\,dz.$$
Given a particular schedule, $g(z)$, we choose the $\alpha$ that
minimizes the above expression to find the minimum average completion
time.  If we wish to restrict ourselves to finding the best schedule
of makespan no more than $1+ \rho$, then we allow $\alpha$ to range
from $0$ to $\rho$, and choose the worst possible schedule $g(z)$.
This corresponds to evaluating
                                            
$$
\max_{g} \min_{0 \leq \alpha \leq \rho}
\int_{\alpha L}^{\infty} \frac{(1 + \alpha) L - z}{z} g(z)\,dz,
$$
where $g$ is a probability distribution over $[0,\infty)$.  This can
be shown to be equivalent to the expression

\begin{equation}
\label{eq:cne}
\max_{f} \min_{0 \leq \alpha \leq \rho}
\int_{\alpha}^{\infty} \frac{1 + \alpha - x}{x} f(x)\,dx,
\end{equation}

where $f$ now ranges over all distributions.

We can show that the maximum of (\ref{eq:cne}) is achieved
by the following function  
\[ 
\fopt(x) = 
\cases{\frac{e^{\rho}}{e^{\rho}-1} x e^{-x} & $0 \leq x < \rho$ \cr
       \frac{\rho}{e^{\rho}-1} \delta(0)    & $x = \rho$        \cr
       0                                    & $x > \rho$.       \cr}
\]
which yields the following bound.

\begin{theorem}
For any $\rho \in [0,1]$, for any scheduling problem, there exists
a $(1 + \rho, e^{\rho}/ (e^{\rho} -1))$ approximation.
\end{theorem}

We omit the proof, but note  that 
this theorem can be verified by viewing the integral as a continuous
infinite-dimensional linear program and computing the dual, which is 

$$\min_h \max_{x \geq 0} \int_0^{\min{\rho,1}}
\frac{(1+\alpha-x)}{x} h(\alpha) d\alpha,$$
where $h$ is a pdf over the interval $ [0,1]$.
This dual is optimized by choosing $h(x) =  e^x/(e^{\delta}-1)$ for
$x \in [0,1]$ and $0$ otherwise.

\begin{corollary}
For any scheduling problem, there exists a $(2, 1.582)$-schedule, a
$(1.695,2)$-schedule and a $(1.806,1.806)$-schedule.
\end{corollary}

These results, for some scheduling models, provide better bicriteria
algorithms than can be achieved by Chakrabarti
et. al.\cite{ChakrabartiPSSSW96} or  Stein and Wein\cite{SteinW97}.
Consider the
case of $w_j=1$ for all $j$. For the scheduling of jobs on
parallel machines of different speeds, there is a polynomial-approximation
scheme for makespan \cite{HochbaumS88}, and a polynomial-time algorithm
for average
completion time\cite{Horn73, BrunoCS74}; resulting in a $(1.806 +
\epsilon,1.806)$-algorithm.  When considering the problem of scheduling jobs on
{\em unrelated} parallel machines 
there is a $2$-approximation algorithm for makespan \cite{LenstraST90}
and a polynomial-time algorithm for average completion time
\cite{Horn73,BrunoCS74}; thus a $(3.612,1.806)$-algorithm exists.
Finally, for the scheduling
of jobs, now with general weights, on parallel identical machines, for average
weighted completion time 
there is a $(\frac{\sqrt{2}+1}{2})$-approximation algorithm
\cite{KawaguchiK86}; together with the polynomial-approximation scheme
for makespan\cite{HochbaumS88} we achieve
a $(1.806 + \epsilon, 2.180)$-algorithm. 

Our results also apply to bicriteria optimization of the travelling
salesman and travelling repairman problems.  In the travelling
repairman problem, we have a start vertex $v$, and define
$c_i$ to be the distance in the tour from vertex $c$ to vertex $i$.
Associated with vertex $i$ is a nonnegative weight $w_i$ and the goal
is to find a tour that minimizes $\sum_i w_ic_i$.  Combining results
from \cite{SteinW97} with the techniques in this paper we see that
the existence of an $(\alpha,\beta)$ schedule implies the existence of
a tour that is simultaneously a $1+\alpha$ approximation for the travelling
salesman problem and a
$\beta$-approximation for the travelling repairman problems.

This model also bounds the completion time of {\em each job}
by a factor of $\beta$ times its completion in an optimal
schedule. Therefore our results have consequences for minsum
criteria other than $\sum w_jC_j$, such as $\sum w_jC^2_j$.

\bibliographystyle{plain}
\bibliography{/u/cliff/latex/bibliography}
\end{document}